\documentstyle[aps,twocolumn]{revtex}

\begin{document}
\title{Tunable Charge Density Wave Transport in a Current-Effect Transistor}
\author{N. Markovi\'{c}, M. A. H. Dohmen, and H. S. J. van der Zant}
\address{Department of Applied Physics and DIMES, Delft University of Technology,\\
Lorentzweg 1, 2628 CJ Delft,\\
The Netherlands}
\date{June 29, 1999}
\maketitle

\begin{abstract}
The collective charge density wave (CDW) conduction is modulated by a
transverse single-particle current in a transistor-like device.
Nonequilibrium conditions in this geometry lead to an exponential reduction
of the depinning threshold, allowing the CDWs to slide for much lower bias
fields. The results are in excellent agreement with a recently proposed
dynamical model in which ''wrinkles'' in the CDW wavefronts are ''ironed''
by the transverse current. The experiment might have important implications
for other driven periodic media, such as moving vortex lattices or ''striped
phases'' in high-T$_{c}$ superconductors.
\end{abstract}

\draft



The charge-density-wave (CDW) state, characterized by a periodic modulation
of the conduction electron density, is commonly observed in low-dimensional
conductors \cite{Peierls}. It is found to be the ground state in various
inorganic and organic materials with a chain-like structure, giving rise to
remarkable electrical properties \cite{Gruner1,Gruner2,Thorne}. Similar
charge-ordered states (''striped phases'') play an important role in high-T$%
_{c}$ superconductors \cite{Kivelson} and two-dimensional electron gases in
the quantum Hall regime \cite{Koulakov}.

A particularly interesting feature of the CDW state is its collective
transport mode, very similar to superconductivity \cite{Frohlich}: under an
applied electric field, the CDWs slide along the crystal, giving rise to a
strongly nonlinear conductivity. Since even a small amount of disorder pins
the CDWs, sliding occurs only when the applied electric field exceeds a
certain threshold field. The pinning mechanisms, the onset of collective
motion and the dynamics of a moving CDW are typical characteristics of the
complex physics which describes a very general class of disordered periodic
media \cite
{Fukuyama,Bardeen,Sneddon,Scheidl,Giamarchi,Balents1,Balents2,Vinokur}.
These include a wide variety of periodic systems, as diverse as vortex
lattices in superconductors and Josephson junction arrays \cite
{Blatter,Reichhardt,Troyanovski}, Wigner crystals \cite{Cha}, colloids \cite
{Chakrabarti}, magnetic bubble arrays \cite{Hu} and models of mechanical
friction \cite{Braun}.

The focus of recent theoretical and experimental research on disordered
periodic media have been their {\it nonequilibrium} dynamical properties.
One of the issues that have been raised is the effect of a single-particle
current, due to uncondensed electrons and quasiparticle excitations. In a
recent theoretical work, Radzihovsky and Toner \cite{Radzihovsky} discovered
that a single-particle current has the most profound effects when it flows
perpendicular to the CDW sliding direction. Based on general symmetry
principles, this leads to nonequilibrium CDW dynamics even if CDW itself is
stationary. Here we report our study of the CDW transport in the presence of
such a transverse single-particle current. We find that the sliding CDW
motion is stable against a small transverse current, but large currents have
a dramatic effect: the longitudinal depinning threshold field is
exponentially reduced for the normal current densities which exceed some
crossover value J$_{c}$. In other words, the collective longitudinal current
is {\it enhanced} by the transverse single-particle current. The
characteristics of this {\it current-effect} transistor are in excellent
agreement with the predictions of Radzihovsky and Toner \cite{Radzihovsky}.

The experiments were carried out on single crystals of NbSe$_{3}$. This
material has a very anisotropic, chain-like structure \cite{Gruner1}. It
exhibits two CDW transitions, each involving different types of chains, at $%
T_{P}=145K$ and $T_{P}=59K$. A small portion of the conduction electrons
remains uncondensed, providing a metallic single-particle channel.

A single crystal of dimensions 2.7 mm$\times $36 $\mu $m$\times $240 nm was
glued onto a sapphire substrate. A pattern of gold contacts was then defined
on top of it using electron-beam lithography. The pattern consisted of two
current leads at two ends of the crystal, and a row of devices, each with
two transverse current leads and two voltage leads. A scheme of such a
transistor device is shown in the inset of Fig. \ref{IV}.

The transverse current leads were 5-100 $\mu $m wide, and overlapped the
crystal by 1-5 $\mu $m. To ensure contact on both sides of the crystal, 180
nm thick layer of gold was evaporated at angles of 45 degrees with respect
to the substrate, as well as perpendicular to it. The contact resistance of
the transverse leads was by 1-2 orders of magnitude larger than the
resistance of the crystal in the longitudinal direction, which precludes
considerable shunting of the current through the transverse leads. A dc
current of up to 1 mA was injected at the transverse leads. The transverse
leads were not electrically connected to the longitudinal circuit, except
through the crystal. Since the CDWs can only slide in the longitudinal
direction, the transverse current is due to single electrons.

The longitudinal current was injected at the two far ends of the crystal.
The voltage leads were 180 nm thick, 5 $\mu $m wide and the spacing between
them was 50-500 $\mu $m. The longitudinal current-voltage characteristics
and the differential resistance were studied as a function of transverse
current at different temperatures, ranging from 25-120 K.

The current-voltage characteristics for one of the devices are shown on Fig. 
\ref{IV}. In the absence of the transverse current, the CDWs are pinned at
low bias voltages. The I-V is linear, as the current is due to uncondensed
electrons and quasiparticles that are thermally excited above the CDW gap.
When the applied voltage reaches the threshold value $V_{T}(I_{x}=0)$,
marked by an arrow in Fig. \ref{IV}, the CDWs are depinned and start to
slide. A sharp increase in current is observed at $V_{T}$ due to this
additional conduction channel.

When a transverse current I$_{x}$ is applied, $V_{T}$ decreases and the
sliding starts at lower bias voltages. Thus, CDWs that were pinned for $%
J_{x}=0$, start sliding at lower fields when a transverse current is
applied. A new linear regime sets in at low bias voltages, where $V_{T}$ $%
<V_{T}(I_{x}=0)$. The resistance in this regime is lower than the single
particle contribution $R$ at $J_{x}=0$. This makes the effect easily
distinguishable from heating: since most of the measurements were carried
out at the temperatures at which $dR/dT>0$, heating would result in a higher
single-particle resistance.

The threshold field reduction is more strikingly visible in the differential
resistance measurements, shown in Fig. \ref{dV/dI}. The differential
resistance at low bias fields, due to uncondensed electrons and excited
quasiparticles, is mostly unaffected by the transverse current. The onset of
CDW sliding, characterized by a sharp drop in differential resistance, is
shifted towards zero as I$_{x}$ is increased. The same reduction of the
threshold field is also observed for negative bias voltages and the plots
are nearly symmetric around $V=0$. We have found no differences when
changing the sign of either the longitudinal current or the transverse
current.

The reduction of the sliding threshold does not occur for arbitrarily small
transverse currents. The dependence of the threshold field $E_{T}$ on the
transverse current density $J_{x}$ for two samples \cite{anisotropy} is
shown in Fig. \ref{EvsJ}. It is evident that $E_{T}$ remains unchanged until 
$J_{x}$ reaches some crossover value $J_{c}$. For $J_{x}>J_{c}$, $E_{T}$
decreases with increasing $J_{c}$. The transverse current density dependence
of the threshold field $E_{T}$ for $J_{x}>J_{c}$ can be fit by:

\begin{equation}
E_{T}(J_{x})=E_{T}(0)\frac{J_{x}}{J_{c}}\exp \left( 1-\frac{J_{x}}{J_{c}}%
\right)  \label{Et}
\end{equation}
where $E_{T}(0)$ is the threshold field at $J_{x}=0$. Once the crossover
value of the transverse current $J_{c}$ is exceeded, the depinning threshold
field decreases and the CDW conduction channel is activated by much lower
bias voltages.

The observation of a crossover current $J_{c}$ rules out the possibility
that the threshold field reduction is due to current inhomogeneities around
the transverse contacts. If the changes in $E_{T}$ were due to a
longitudinal component of an inhomogeneous transverse current, then such
changes would be apparent at any value of $J_{x}$, and no $J_{c}$ would be
observed. Furthermore, it is not clear that such inhomogeneities would lead
to the observed exponential reduction of $E_{T}$.

The exponential decrease of the threshold field described by Eq. \ref{Et}.
has recently been predicted by Radzihovsky and Toner \cite{Radzihovsky}. In
their model, the value of the crossover current density $J_{c}$ needed for
the initial suppression of $E_{_{T}}$ is expected to be proportional to the
value of the threshold field at $J_{x}=0$, and is given by \cite{Radzihovsky}%
:

\begin{equation}
J_{c}\propto \sigma _{0}E_{T}(0)\;(\xi _{L}k_{F})\;(\rho _{n}/\rho _{CDW})
\label{Jc}
\end{equation}
where $\sigma _{0}$ is the conductivity at very high bias fields, $k_{F}$
the Fermi wave vector and $\rho _{n}$ and $\rho _{CDW}$ are normal and CDW
electron densities, respectively. The correlation length $\xi _{L}$ \cite
{Fukuyama,Larkin} is a measure for the coherence in the sample and decreases
with increasing disorder.

$E_{T}$ is known to be temperature dependent, following $%
E_{T}=E_{T}(0)e^{-T/T_{0}}$ \cite{McCarten}, where $T$ is the temperature,
and $T_{0}$ is a constant. The dependence of $J_{c}$ on $E_{T}(0)$ can
therefore be studied by measuring at different temperatures. The dependence
of $J_{c}$ on $\sigma _{0}E_{T}(0)$ is shown in the inset of Fig. \ref{EvsJ}%
: J$_{c}$ grows linearly with $\sigma _{0}E_{T}(0)$ and it extrapolates to
zero for $E_{T}(0)=0$. The crossover current densities of $10^{3}-10^{4}$
A/cm$^{2}$ estimated from Eq. \ref{Jc}. \cite{Radzihovsky} are in excellent
agreement with the values measured in our experiment.

We have shown that the conduction in the CDW channel can be enhanced by a
single-particle current flowing transversely to the CDW sliding direction.
This surprising behavior has been observed in samples with different
geometries, at different temperatures, and in both CDW regimes of NbSe$_{3}$%
, suggesting that it is a general property of the CDW transport.

The dynamical model of Radzihovsky and Toner \cite{Radzihovsky} provides a
physical origin of this effect: the CDWs become more ordered due to momentum
transfer with transversely moving normal carriers. This mechanism is
illustrated in Fig. \ref{Model}. In the absence of defects, the charge
density wave fronts are straight and parallel to each other (left side of
the picture). The single-particle transverse current, marked by ''a'' on
Fig. \ref{Model}, can flow with little or no interaction with the CDW. In
the presence of defects or impurities in the crystal, the CDW deforms to
lower its energy and the wavefronts are ''wrinkled'' (right side of the
picture). In this case, the transversely moving electrons (''b'') are more
likely to be deflected. The conservation of linear momentum results in a
reaction force back on the CDW. This way the CDW roughness is reduced as the
CDW wavefronts are straightened out or ''ironed'' by the transverse current.
The CDW transport across the sample is therefore more coherent and less
susceptible to pinning. The lower pinning strength then leads to a lower
threshold field.

Since the conduction in the CDW channel can be modulated by a current in the
single particle channel, this device in principle works as a transistor,
raising a question of a possible practical application. The maximum gain
observed in our experiments was $\Delta I/I_{x}=0.15$. A simple estimate
from our measurements suggests that the maximum gain is proportional to $\xi
_{L}^{-1}$. The gain can therefore be improved by using dirtier crystals or
smaller samples in which $\xi _{L}$ is limited by the sample sizes.

Apart from being intriguing in their own right as an important test of the
theory, our results may provide a useful insight into related phenomena
which are much more difficult to study experimentally. As mentioned above,
this novel effect is relevant to a variety of other periodic systems which
share the same symmetries and a similar geometry. A particularly interesting
example might be the ''striped phases'' in superconducting oxides, whose
role in high-T$_{c}$ superconductivity is still not resolved.

The authors are grateful to Yu. Latyshev and P. Monceau for providing the
crystal, and to L. Radzihovsky, Yu. Nazarov and S. Zaitsev-Zotov for useful
discussions. This work was supported by the Netherlands Foundation for
Fundamental Research on Matter (FOM). HSJvdZ was supported by the Dutch
Royal Academy of Arts and Sciences (KNAW).

%

\begin{figure}[tbp]
\caption{Current-voltage characteristics for a 36 $\mu $m wide and 0.24 $\mu 
$m thick NbSe$_{3}$ crystal at 45 K with the values of the transverse
current ranging from 0 (bottom) up to 800 $\mu $A (top). The width of the
transverse current leads was 100 $\mu $m, and the spacing between the
voltage leads was 225 $\mu $m. The dotted line represents the Ohmic behavior
of the uncondensed electrons. The data deviate from that line when the
charge density waves are depinned and start to slide, contributing
additional current. The depinning treshold V$_{T}$, marked by an arrow,
decreases when a transverse current I$_{x}$ is applied. Inset: The scheme of
the transistor-like device, which consists of the crystal (dark shaded area)
and six gold leads (light shaded areas): I for the longitudinal current,
which is injected at the far ends of the crystal, V for voltage measurements
and I$_{x}$ for the transverse current injection.}
\label{IV}
\end{figure}

\begin{figure}[tbp]
\caption{Differential resistance as a function of electric field at 45 K of
the same device as in Fig.~1. Different symbols represent different values
of the transverse current I$_{x}$, ranging from 0 (filled circles) to 190$%
\mu A$ (filled squares). The treshold field at which the differential
resistance drops due to the depinning of the charge density waves is shifted
to lower values as the transverse current is increased.}
\label{dV/dI}
\end{figure}

\begin{figure}[tbp]
\caption{Depinning treshold field E$_{T}$, scaled by its value at J$_{x}$=0,
as a function of transverse current density J$_{x}$ at 55 K for two
different devices. The widths of the transverse current leads were 100 $\mu
m $ (circles) and 30 $\mu m$ (triangles). The solid line represents a fit to
Eq. 1. Inset: The crossover transverse current density J$_{c}$ needed for
the initial supression of the depinning treshold E$_{T}$ increases with the
value of E$_{T}$ at J$_{x}$=0. J$_{c}$ is plotted here as a function of $%
\sigma _{0}E_{T}(0)$ for comparison with Eq. 2, where $\sigma _{0}$ is the
conductance at bias fields much larger than $E_{T}$. Each data point was
determined at a different temperature: from right to left, the temperatures
were 25, 30, 35, 40, 45, 50 and 55 K.}
\label{EvsJ}
\end{figure}

\begin{figure}[tbp]
\caption{Dynamical model for the threshold field reduction: the vertical
lines represent the charge density wavefronts which, if depinned, can move
in the horizontal direction. The large filled circles represent the
impurities or other defects in the crystal. The CDW sliding direction and
the transverse current direction are indicated by the arrows. The small open
circles (''a'' and ''b'') represent the transverse single particle current.
The deformations of the wavefronts due to momentum transfer with the
transverse current are shown by the dotted lines.}
\label{Model}
\end{figure}


\end{document}